# Interplay of $Sm^{4f}$ and $Co^{3d}$ spins in SmCoAsO

V.P.S. Awana[*], Anand Pal, Bhasker Gahtori, and H. Kishan

Quantum phenomenon and applications (QPA) division, National Physical Laboratory (CSIR),

Dr. K. S. Krishnan Marg New Delhi -110012, India

We present detailed magnetization and magneto-transport studies on the title compound SmCoAsO. In a recent paper we reported [1] the complex magnetism of this compound. SmCoAsO undergoes successive paramagnetic (*PM*) – ferro-magnetic (*FM*) – anti-ferro-magnetic (*AFM*) transitions with decrease in temperature. This is mainly driven via the c-direction interaction of $Sm^{4f}$ (SmO layer) spins with adjacent (CoAs layer) ordered $Co^{3d}$ spins. In this article we present an evidence of kinetic arrest for *FM-AFM* transition. The isothermal magnetization (*MH*) loops for SmCoAsO exhibited the meta-magnetic transitions at 6, 8 and 10K at around 80, 60 and 50kOe fields respectively with characteristic hysteresis shoulders along with the non-zero moment at origin, thus suggesting the possibility of kinetic arrest. Suggested kinetic arrest is further evident in zero field-cooled (*ZFC*) and field-cooled (*FC*) hysteresis under high fields of up to 140kOe magnetization (*MT*) and the magneto-transport measurements *R(T)H* during *FM-AFM* transition. The time dependent moment experiments exhibited very small (~2-3%) increase of the same below 20kO and decrease for 30kOe at 15K.

[*]Corresponding author:

Dr. V.P.S. Awana

Quantum Phenomenon and Application Division,

National Physical Laboratory (CSIR), New Delhi-110012, India

E-mail: awana@mail.nplindia.ernet.in

Web page: www.freewebs.com/vpsawana/

Introduction

Since the recent discovery [2] of superconductivity in doped REFeAsO (RE= rare earths) oxy-pnictides, similar structure cobaltates (RECoAsO) have attracted renewed interest [3-5]. Co orders ferro-magnetically (*FM*) below 100K in case of LaCoAsO with saturation moment of 0.10 – 0.20 $\mu_B$B/Co below Curie temperature ($T_c$) and relatively higher effective paramagnetic moment of above 1.50 $\mu_B$/Co above $T_c$, thus indicating an itinerant *FM* state [1, 3-5]. The situation becomes more interesting and rather complex with gradual *FM-AFM* transitions, when the non-magnetic La is replaced with magnetic Nd [6, 7], Sm [1, 5, 8] or Ce [5,9]. It seems that magnetic RE spins influences the *FM* ordering of Co sub-lattice and converts the same to *AFM* at lower temperatures. In such situations a natural question arises as if the *FM-AFM* state is kinetically arrested or not. This has been evidenced in number of well experimented magnetization studies on say CeFe$_2$ [10, 11] and some magnetically phase separated manganites [12], where an *FM* to *AFM* transformation takes place gradually at low temperature. In this article we address the kinetic arrest of *FM* to *AFM* transition in SmCoAsO and compliment our earlier recent work on same compound [1]. Our high field magnetization and magneto-transport studies establish the *FM-AFM* kinetic arrest in SmCoAsO for first time.

Experimental

Polycrystalline SmCoAsO is synthesized by single step solid-state reaction method via vacuum encapsulation technique [1]. Stoichiometric amounts of high purity (~99.9%) Sm, As, Co$_3$O$_4$ and Co powders are ground thoroughly using mortar and pestle. The mixed powders were palletized and vacuum-sealed ($10^{-4}$ Torr) in a quartz tube. These sealed quartz ampoule was placed in box furnace and heat treated at 550$^o$C for 12 hours, 850$^o$C for 12 hours and then finally 1150$^o$C for 33 hours in continuum with slow heating rate, details of synthesis are further given in ref. 1. The entire heating process is "Single Step" in nature as reported earlier by some of us [13]. The X-ray diffraction pattern of the compound was taken on Rigaku X-ray diffractometer with Cu K$_\alpha$ radiation. The resistivity measurements were carried out by conventional four-probe method on *PPMS* (Physical Property Measurement System) from Quantum Design -USA. Heat capacity and magnetization measurements were also carried out on the same *PPMS*.

Results and Discussion

As synthesized SmCoAsO is crystallized in near single phase with lattice parameters $a$=3.957(3)Å and $c$=8.242(2)Å [1]. Magnetic moments (*M*) versus temperature (*T*) plots for SmCoAsO in field-cooled situation at varying applied fields from 10Oe to 90kOe are depicted in Figure 1. The compound undergoes *PM-FM-AFM* transitions with lowering of temperature. Also at low fields of 10 Oe it exhibits two peaks at around 57 and 45K [1]. For higher fields of above 100Oe only one single peak for *FM-AFM* transition is evident. With increase in applied field the *FM-AFM* transition peak temperature decreases to below 10K for 100kOe field. This is in accord with earlier reports on similar compounds [1, 5, 8].

To check if the *FM* is kinetically arrested during the *FM-AFM* transition; we carried out the magnetization measurements in both field-cooled (*FC*) and zero-field-cooled (*ZFC*) situation in higher fields of up to 140kOe. These results are depicted in Figure 2. For an applied field of 30kOe there is nearly no branching of *FC* and *ZFC*. However as the field is increased to 60, 80, 120 and 140kOe the *ZFC* and *FC* branching is clearly evident at 11, 9, 7 and 6K respectively. The *FC* and *ZFC* branching seen at 140kOe field cannot be due to spin-glass (SG) magnetic phase [11, 12]. At this juncture we tend to believe that possibly the *FM* phase is kinetically arrested during the *FM-AFM* transition in SmCoAsO. Also it is evident from Figure 2 that the kinetic arrest of *FM* phase if at all present is for higher fields i.e. > 30kOe. To further establish the possible kinetic arrest, we cooled the sample under 140kOe field from 250K i.e., from within paramagnetic state and carried out the *MT* down to 3K and the isothermal magnetization (*MH*) is recorded from the cooling field 140kOe to zero and checked if there is any non-zero moment left at the origin. We found that the compound bears a non-zero moment at origin. The high field isothermal magnetization (*MH*) results for SmCoAsO will be discussed later.

Relatively lower field (< 30kOe) isothermal magnetization (*MH*) results for SmCoAsO are presented in Figure 3 for various temperatures in *PM*, *FM* and *AFM* regions. The *MH* is linear in both paramagnetic (150K) and anti-ferrmagnetic (2K, 5K) regions. For *FM* regions (40, 60 and 80K) the same is *FM* like with saturation moment of ~ 0.20 $\mu_B$. This is similar to that as reported earlier for similar compounds [6-9]. The *MH* plots exhibit clear meta-magnetic transitions at 10K and 20K respectively at 10kOe and 20kOe fields. These plots are marked in Figure 3. Also the MH plots at 20K exhibit characteristic shoulder hysteresis at 20kOe field. Interestingly these temperatures of 10K and 20K fall in the region where the compound is in the middle of *FM-AFM* transformation. Clearly in this region the competing *FM* and *AFM* states give rise to meta-magnetic instability, which is being reminiscent in the *MH* plots in Figure 3.

High field (120kOe) field isothermal magnetization (*MH*) plots for SmCoAsO at 4, 6, 8 and 10K are depicted in Figure 4. This is precisely to check possible reminisce of *FM* clustering with in *AFM* dominated phase or the kinetic arrest of *FM*. The *MH* plot at 4K is nearly linear without any meta-magnetic deviation or visible hysteresis. The *MH* plot at 6K is also though nearly linear but with hysteresis opening between 40 to 120kOe. Both the 8 and 10 K *MH* plots exhibit shallow/rounded meta-magnetic like steps at around 80 and 60kOe respectively with hysteresis at these shoulders. These hysteresis shoulders are similar to that as observed at 20K for low field (< 30kOe) *MH* data shown in Figure 3. The meta-magnetic hysteresis shoulders shown at 20K, 20kOe field in Figure 3 and at 8K and 10K for 80 and 60kOe fields imply for the complex magnetic interactions in the compound. This is unlike LaCoAsO, where the $Co^{3d}$ spins order *FM* below 100K, and remain the same down to low temperatures [1, 3-5]. In present case the $Sm^{4f}$ and $Co^{3d}$ spins interplay with each other and thus the *FM* ordered $Co^{3d}$ spins convert to *AFM* arrangement with lowering of temperature and changing field. It is the competing effect of $Sm^{4f}$ and $Co^{3d}$ spins interplay which gives rise to meta-magnetic state in intermediate range of temperature and field. These temperatures and fields correspond to the mixed state whereby the *FM* state is not converted fully into *AFM*. To check if the *FM* kinetic arrest is taking place at low temperatures, the isothermal magnetization (*MH*) loops are recorded from higher fields (120kOe) and the date is zoomed. The inset of Figure 4 shows the non-zero moment at origin at temperatures of up to 10K. The non-zero moment at origin calls for possible kinetic arrest of *FM* for these higher fields.

Resistance versus temperature *R(T)* plots in various applied fields of at 0, 5kOe, 10kOe, 50kOe and 140kOe fields are depicted in Figure 5. *R(T)* of SmCoAsO is metallic in studied temperature range of 300-2K. The metallic slope remains invariant down to 100K, and increases due to onset of *FM* state at this temperature, which decreases below 50K due to the presence of the competing *AFM* phase. Hence in general though the *R(T)* of SmCoAsO remains metallic through the studied temperature range of 300K down to 2K, the *change in metallic slope* takes place corresponding to *PM-FM-AFM* magnetic transitions. We also carried out the magneto-transport *R(T)H* measurements at various fields. Magneto-resistance is clearly seen below *FM* transition (100K). The deviation of resistance under magnetic field though increases with lowering of temperature, a step appears in *R(T)H* at a particular field and temperature bringing the resistivity to zero field value with a *step like* transition. This strictly means the magneto-resistance practically comes down to zero at a particular temperature and field. This is though qualitatively similar behavior to that as reported for NdCoAsO [6], the important difference is that no step is seen in zero field *R(T)H* of SmCoAsO. Further after applying field the step like up-turn in *R(T)H* brings the *R* back to the zero field value. These steps are more clearly depicted in inset of Fig. 5. Namely these *R(T)H* steps occur at 40K, 32K, 18K and 8K respectively for 5kOe, 10kOe, 50kOe and 140kOe applied fields, see inset Figure 5. Interestingly, the bringing back of the resistivity to zero field value with a *step like* transition is complete for all fields except for 140kOe, see inset Figure 5. This means when the sample is cooled under 140kOe field, the *FM* phase is kinetically arrested and its full conversion to *AFM* does not take place. Worth noting is the fact, that these step temperatures lies in the region where SmCoAsO undergoes through *FM-AFM* transformation region, i.e. the conversion of *FM* to *AFM* being not complete. Although one could think of these temperatures for unusual ordering of $Sm^{4f}$ spins, but that is highly unlikely. Further, the conduction process in SmCoAsO is through CoAs layers only, and hence the *R(T)H* do represent mainly the contribution from ordered $Co^{3d}$ spins. Although the ordering of $Co^{3d}$ spins is influenced by $Sm^{4f}$ spins, but the later has no direct role in conduction process or the spin scattering.

To elucidate upon the *R(T)H* results of Figure 5 more clearly, the fixed temperature (200K, 100K, 50K, 20K, 10K, 5K and 2.5K) and varying field (±100 kOe) isothermal magneto-resistance (*MR%*) results are shown in Figure 6. The *MR%* is defined as $[(R_H-R_o)/R_H]*100$. At 200K, i.e., in *PM* state the compound exhibits small +ve *MR*, which is usual. As the temperature is lowered to 100K, the compound enters into *FM* state and exhibits –*MR* of up to 6% at 100kOe, and the shape of the MR plot is nearly linear in both increasing/decreasing field. At 50K, the –ve *MR* is further increased to around 12% albeit with non-monotonic variation with field. The same shows sharp increase at low fields (<20kOe) and relatively smaller increase for higher fields. Interestingly 50K is the temperature where SmCoAsO is already started competing for *FM-AFM* transformation. At low fields the majority phase is *FM* with higher *LFMR* (low field magneto-resistance). On the other hand at higher fields the *AFM* starts dominating and hence a decreasing rate of change in *MR*. However the overall *MR* at 100kOe and 50K is yet double to that of at 100K and same field, which is mostly the *FM* phase alone. This increase of *MR* at 50K may be due to the magnetic phase separation [14] of the two competing phases (*FM/AFM*) being present at these temperatures. With further lowering of temperature to 20K, the maximum *MR* increases to 16% at 100kOe field, with U type shape. This temperature of 20K is just in the middle of *FM-AFM* transition. The *MR* is nearly saturated at around 16% above 30kOe. This basically means that above 30kOe the increasing *FM* contribution to *MR* with field is nearly compensated by the growing *AFM* phase and hence *MR* is nearly saturated. At 10K, the situation is even more interesting at this temperature the *AFM* phase is dominating and hence the over all –ve *MR* is decreased from 20K value at all fields. Further the shoulder hysteresis is seen at around 60kOe field while increasing/decreasing the field in both cycles. This is marked with arrows in Figure 6. This reminds of the meta-magnetic transitions being seen in magnetization of this sample in Figure 3. With still lower temperatures of 5, and 2.5K the *MR* is yet negative but with reduced values of >3% only. This means that *FM* contribution (mainly responsible for –ve *MR*) is though yet present at these low temperatures but is nearly taken over by the *AFM*. The small –ve *MR* of SmCoAsO at 2.5 and 5K yet again approve the possible arrest of *FM* phase in this system, even at low temperatures and higher fields of up to 10kOe. It is clear that the interesting behavior (shape/magnitude) of *MR* being shown in Figure 6 is the result of competing *FM/AFM* phases. In our knowledge this is first study on competing *FM/AFM* magnetic phases and resultant high field and varying temperature *MR* behavior of the SmCoAsO.

Figure 7 depicts the heat capacity (*Cp*) of presently studied SmCoAsO compound. The absolute value of *Cp* at 200K is nearly the same as reported earlier by some of us [1]. The *AFM* ordering of Sm spins is seen clearly at 5K in *Cp(T)* plot, but the Co spins ordering is not seen, suggesting the transition to be possibly of first order. To check upon this we plotted the *Cp/T* vs *T* for SmCoAsO in the inset of Figure 7. This is same as being done in case of CeCoPO [9]. Albeit, the ordering of Co is seen in $C_P(T)$ plot as a broad peak in ref. 9 for CeCoPO. This is not the situation for our studied SmCoAsO, may it be that the closer look of plots at short intervals reveal some thing. Hence this is yet left as an open question. As of now, what is clear, that only 5K ($T_N$, Sm) is visible in *Cp(T)* measurements on presently studied SmCoAsO. This is unlike NdCoAsO, for which a clear but broad peak is seen at around 15K for $T_N$ of Nd [6]. We also carried out the magneto heat capacity of the compound at 100kOe field and the result is shown in upper inset of

Figure 7. It is clear that the $C_P(T)H$ is deviated from the zero field value at low temperatures below 20K. This is unlike the case of SmFeAsO, where there is nearly no change in $C_P(T)H$ under magnetic field [15]. This is precisely because unlike in present case of SmCoAsO, in SmFeAsO the $Sm^{4f}$ and $Fe^{3d}$ spins do not interplay, and their respective ordering are seen distinctly at different temperatures. In present case the interacting $Sm^{4f}$ and $Co^{3d}$ spins alters ordered $Co^{3d}$ spins *FM* to lower temperatures due to competing *AFM*. As the *FM-AFM* competing phenomenon is both temperature and field dependent the change in entropy is seen over a wide temperature range of temperature up to 20K, instead of simply at 5K due to $T_N$ of Sm. The absence of $C_P$ peak at Co ordering temperatures indicate that the transition to magnetic states is of first order and thus possibility of kinetic arrest. But it is not yet conclusive, because for a similar compound (NdCoAsO), a clear but broad peak is seen in $C_P(T)$ at around 15K for unusual $T_N$ of Nd [6]. It is worth mentioning that the interaction of $RE(Sm/Nd)^{4f}$ and $Co^{3d}$ and the resultant kinetic arrest of Co *FM* spins will depend upon the c-lattice parameter, i.e. the z-direction distance between Sm/Nd and Co. This could give rise to different behavior of $Cp(T)$ for NdCoAsO [6] than the one observed for presently studied SmCoAsO.

We also carried out time dependent magnetization experiments in possible kinetically arrested regime i.e., around 15K, these results are shown in Figure 8. The moment (*M*) changes though marginally (~2-3%) with time (*t*), in particular, the same increases for applied field of 5kOe and 20kOe but decreases at 30kOe fields. All the isothermal *M(t)* experiments are carried out at 15K. It is clear that the moment relaxation at 15K is field dependent. This further shows the complex field and temperature dependent of possibly kinetically arrested magnetism of SmCoAsO.

Summarily, we have presented high field magnetization, magneto-transport and heat capacity studies on SmCoAsO compound, and complimented our recently reported [1] results on the same. The intriguing *PM-FM-AFM* transition is probed in detail and evidence is given for the kinetic arrest of *FM* phase during *FM-AFM* transition in this system. The magneto-transport study being carried out for first time on SmCoAsO closely relates with the high field magnetization and changing magnetic structure of the compound during *FM-AFM* transition. The *FM* to *AFM* competition with field and temperature is clearly evident in high field magneto-transport of SmCoAsO.

Authors would like to thank Prof. P. Chaddah Director, CSR-IUC Indore for various discussions related to this work. Authors also acknowledge the keen interest and support of their Director Prof. R.C. Budhani in this work.

Figure Captions:

Figure 1 $M(T)$ down to 5K in field cooled situation at fields from 10 Oe to 90kOe for SmCoAsO.

Figure 2 $M(T)$ in both ZFC and FC situations at 30, 80, 100, 120 and 140kOe fields for SmCoAsO.

Figure 3 $M(H)$ for SmCoAsO till 30kOe applied fields in all five quadrants, meta magnetic field is seen clearly at 20K and around 20kOe with hysteresis in shoulder as marked.

Figure 4 $M(H)$ for SmCoAsO till 120 kOe applied field in all five quadrants, the meta magnetic transitions are seen clearly at 6, 8 and 10K at around 80, 60 and 50kOe fields respectively with characteristic shoulders and hysteresis.

Figure 5 $R(T)H$ for SmCoAsO at 0, 5, 10, 50 and 140kOe fields, inset shows the extended low temperature region.

Figure 6 Magneto resistance $R(H)$ at 200, 100, 50, 20, 10, 5 and 2.5K in applied fields of up to 100kOe during both increasing/decreasing the field.

Figure 7 Heat capacity $C_P(T)$ for SmCoAsO; upper inset shows the extended $C_P(T)$ at 0 and 100kOe fields and lower inset show the $C_P/(T)$ vs $T$ plot for the same.

Figure 8 Moment ($M$) versus time ($t$) plots for SmCoAsO at 15K for 20kOe, 30kOe and 50kOe field

Fig. 1

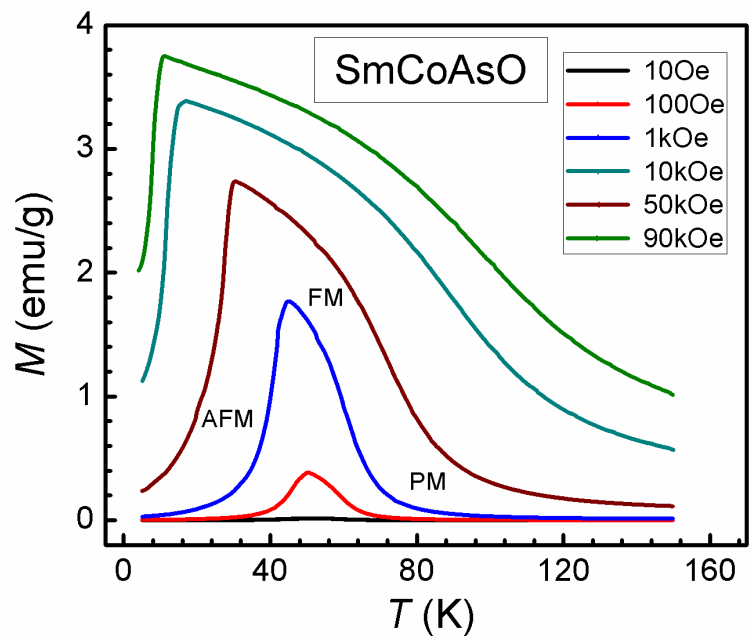

Fig. 2

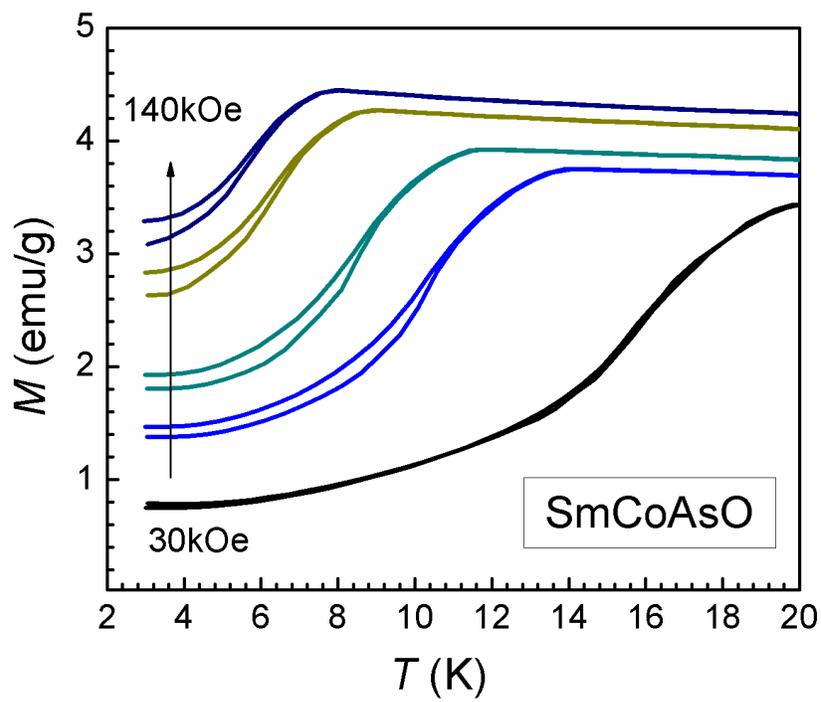

Fig. 3

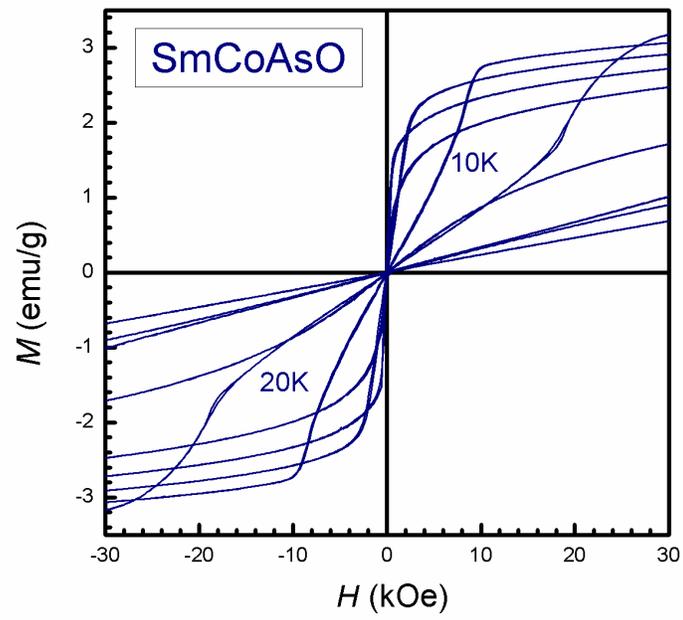

Fig. 4

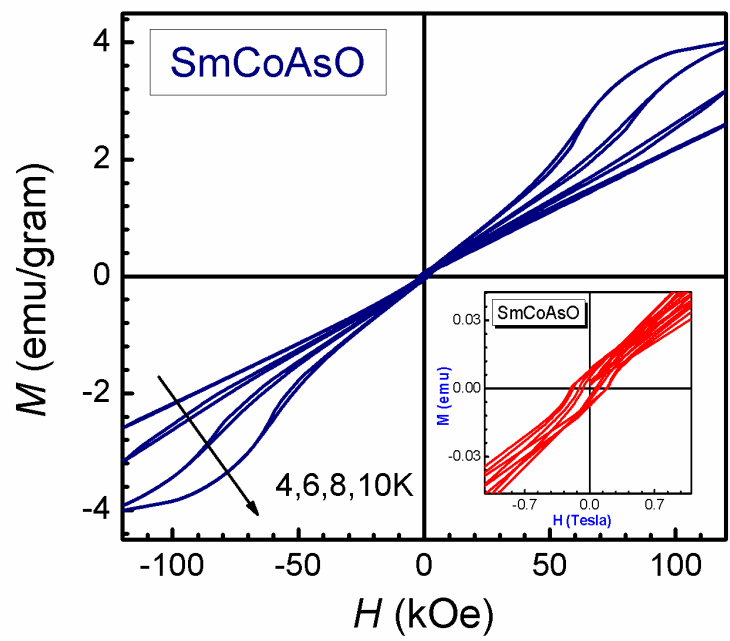

Fig. 5

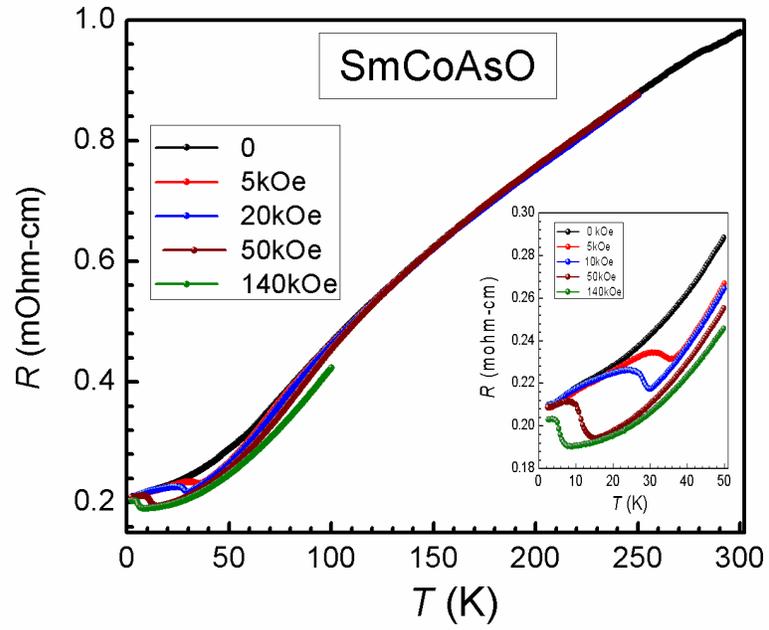

Fig. 6

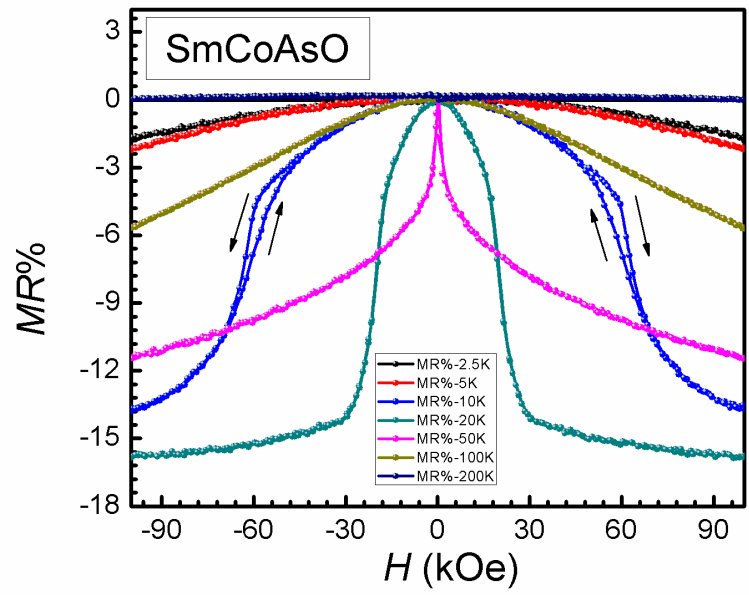

Fig. 7

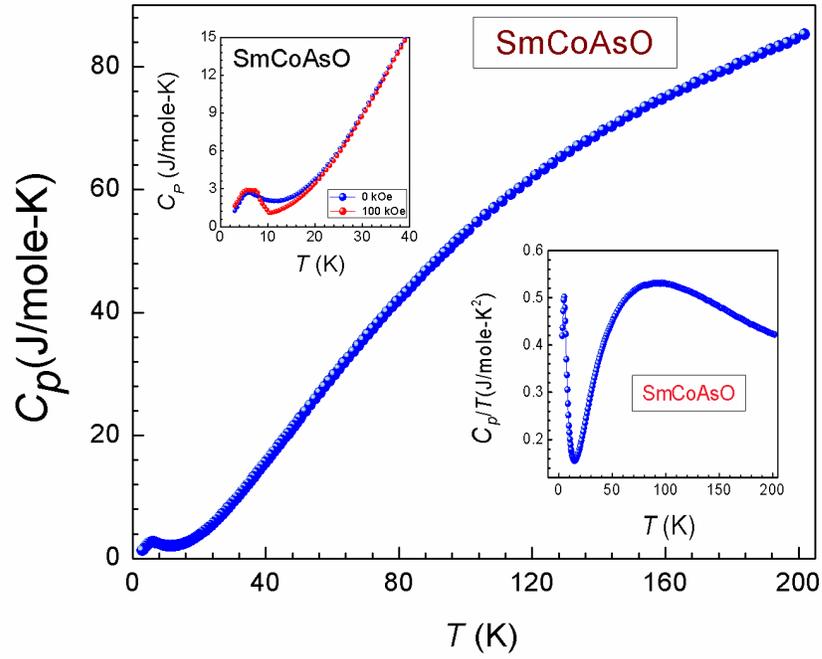

Fig. 8

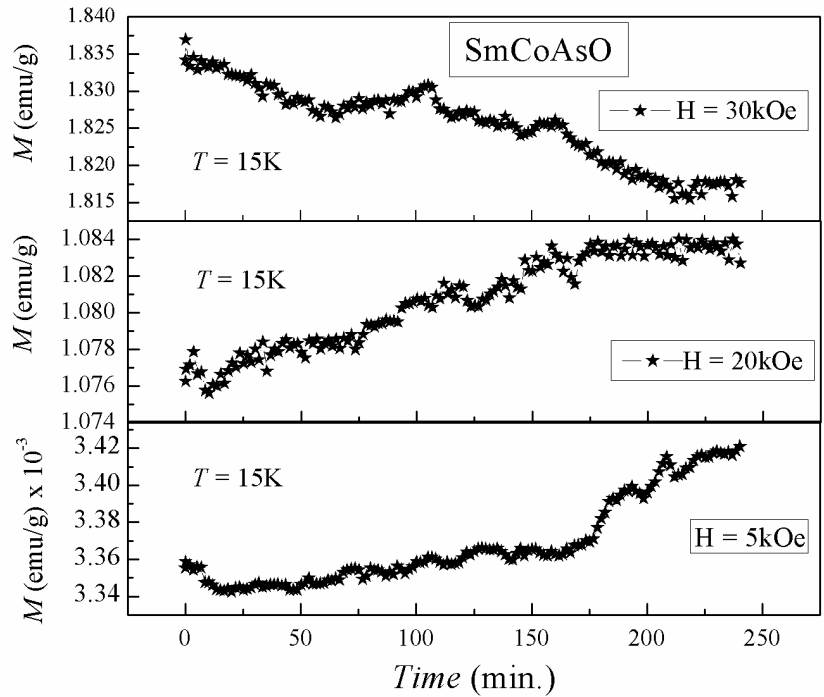

13